# Tangential and quantum frictional forces on a neutral particle moving near a surface

G.V.Dedkov[1] and A.A.Kyasov

Nanoscale Physics Group, Kabardino –Balkarian State University, Nalchik, 360004

Russian Federation

**Abstract**

We calculate tangential forces applied to a ground state atom (nanoparticle) moving with nonrelativistic velocity parallel to the surface of Drude –modelled or Lorentz –modelled half –space using the formalism of fluctuation electrodynamics. The obtained formulae assume the nonretarded regime and arbitrary temperatures of the particle and medium. The quantum frictional force is realized at zero temperature of the particle and medium and can reach about 1/3 of the static value of the Casimir –Polder force in the case of ground state atom. In the case of nanoparticles and nonzero temperature, the tangential forces appears to be both frictional and accelerating. Numerical examples are given for atoms of Na above the surface of Au and MgO nanoparticles above the surface of SiC.

## 1. Introduction

Usually, the so called quantum frictional forces between two perfectly flat parallel dielectric surfaces in relative motion, separated by vacuum gap at zero temperature [1,2], are considered as to be a consequence of fluctuation mechanism, underlying the van der Waals forces of attraction. In general case, it is worthwhile to speak about tangential forces arising between the bodies in relative motion, because the velocity dependence of such forces can be essentially nonlinear and at definite conditions (at non zero temperatures) these forces can even be accelerative [3]. For a long time, this problem has been triggered by several controversies (see, for example [3] for a review of nonrelativistic case) and is being intensively debated to date [2,4-11].

The present paper aims at summarizing our results [3,9] relevant to tangential van der Waals forces between a small particle moving with arbitrary but nonrelativistic velocity and the solid surface. We formulate conditions which are favorable for the particle acceleration and draw attention to the fact that in the case of cold particle and surface, the value of tangential force (decelerating) can be of as large as 1/3 of the value of attractive Casimir –Polder force. We give numerical examples of the quantum frictional force for atoms of Na above the gold surface and accelerative force –for MgO nanoparticle above the heated SiC surface.

As far as concerned the quantum frictional forces at zero temperature of both the particle and surface, our results agree with [1,2] and with recent result of Barton [8] obtained in quantum perturbation theory.

---

[1] Corresponding author e-mail: gv_dedkov@mail.ru

The nonrelativistic limit of tangential force obtained in [10] (eq. (23)) also tallies with ours's [3,9], but if retardation is taken into account, the corresponding results [10,11] proved to have serious drawbacks [9]. Moreover, despite that the authors [10,11] have taken into account the temperature effects in calculations of tangential forces, they disregarded a possibility of accelerating tangential force.

## 2. Theory

Now we recall our nonrelativistic expressions for the tangential force $F_x$. So, following [3],

$$F_x = -\frac{2\hbar}{\pi^2} \int_0^{+\infty} dk_x k_x \int_0^{+\infty} dk_y k \exp(-2kz_0) \int_0^{\infty} d\omega \cdot \left\{ \coth\left(\frac{\hbar\omega}{2k_B T_1}\right) \alpha''(\omega) \cdot \right.$$
$$\left. \cdot [\Delta''(\omega + k_x V) - \Delta''(\omega - k_x V)] + \coth\left(\frac{\hbar\omega}{2k_B T_2}\right) \Delta''(\omega) [\alpha''(\omega + k_x V) - \alpha''(\omega - k_x V)] \right\} \quad (1)$$

where $T_1$ and $T_2$ are the temperatures of the particle and surface, double primed $\alpha''(\omega)$ denotes imaginary part of the particle polarizability,

$$\Delta(\omega) = \frac{\varepsilon(\omega) - 1}{\varepsilon(\omega) + 1}, \quad (2)$$

$\varepsilon(\omega)$ denotes bulk dielectric permittivity of half–space, and $\Delta''(\omega)$ denotes imaginary part of $\Delta(\omega)$. Moreover, $z_0$ is the particle distance from the surface and $V$ is the velocity taken to be parallel to the surface in $x-$ direction.

A more compact form of (1) is obtained introducing new variables $\omega \pm k_x V = \omega'$ in the line which contains $\Delta''(\omega \pm k_x V)$ and replacing limits of integration $0 \leq k_x, k_y < \infty$ by $-\infty \leq k_x, k_y < \infty$. Then, after simple transformations we get

$$F_x = -\frac{\hbar}{\pi^2} \int_{-\infty}^{+\infty} dk_x k_x \int_{-\infty}^{+\infty} dk_y k \exp(-2kz_0) \int_0^{\infty} d\omega \Delta''(\omega) \alpha''(\omega + k_x V) \cdot$$
$$\cdot \left[ \coth\left(\frac{\hbar\omega}{2k_B T_2}\right) - \coth\left(\frac{\hbar(\omega + k_x V)}{2k_B T_1}\right) \right] \quad (3)$$

Just the same form (3) stems from our relativistic expressions for $F_x$ [9] in the limit $c \to \infty$.

Assuming $T_1 = T_2 = 0$ with account of relations

$$\lim_{T_2 \to 0} \coth\frac{\hbar\omega}{2k_B T_2} = sign(\omega)$$

$$\lim_{T_1 \to 0} \coth \frac{\hbar(\omega + k_x V)}{2 k_B T_1} = sign(\omega + k_x V),$$

eq.(3) takes the form

$$F_x = \frac{4\hbar}{\pi^2} \int_0^{+\infty} dk_x k_x \int_0^{+\infty} dk_y k \exp(-2k z_0) \int_0^{k_x V} d\omega \, \Delta''(\omega) \alpha''(\omega - k_x V) \tag{4}$$

Eq.(4) has been reported firstly in Ref. [3b] (eq.(4.30)) and describes quantum frictional force analogous to eq.(12) in Ref. [2] in the case of two weakly interacting half –spaces. By definition of the functions $\alpha(\omega), \Delta(\omega)$, the force (4) is always negative, giving rise to deceleration of the particle. A simple integration (4) over $k_y$ yields

$$F_x = \frac{2\hbar}{\pi^2} \int_0^{+\infty} dk_x k_x^3 \left[ K_0(2 k_x z_0) + K_2(2 k_x z_0) \right] \int_0^{k_x V} d\omega \, \Delta''(\omega) \alpha''(\omega - k_x V) \tag{5}$$

where $K_{0,2}(x)$ are the McDonald's functions. One sees from (5) that dependence $F_x$ on velocity $V$ is essentially nonlinear. The outer integrand function is maximal at $k_x \sim 1/z_0$ and therefore, the characteristic frequencies involved in the inner integrand function, should be of order $V/z_0$.

To analyze possibility of accelerative force which stems from (3), it is worthwhile to consider a nondissipative approximation for the particle and surface dielectric properties:

$$\alpha(\omega) = \frac{\alpha(0) \omega_0^2}{\omega_0^2 - \omega^2 - i \cdot 0 \cdot sign(\omega)} \tag{6}$$

$$\varepsilon(\omega) = 1 - \frac{\omega_P^2}{\omega(\omega + i \cdot 0)} \tag{7}$$

From (6),(7) it follows

$$\alpha''(\omega) = \frac{\pi \alpha(0) \omega_0}{2} \left[ \delta(\omega - \omega_0) - \delta(\omega + \omega_0) \right] \tag{8}$$

$$\Delta''(\omega) = \frac{\pi \omega_s}{2} \left[ \delta(\omega - \omega_s) - \delta(\omega + \omega_s) \right] \tag{9}$$

Substituting (8),(9) into (3) and performing integration over $\omega$ and $k_y$ yields

$$F_x = -\frac{\hbar \alpha(0) \omega_0 \omega_s}{64 z_0^4} \left\{ \frac{1}{(\omega_0 - \omega_s)} x_1^4 \left[ K_0(x_1) + K_2(x_1) \right] \cdot \left[ \coth\left(\frac{\hbar \omega_s}{2 k_B T_2}\right) - \coth\left(\frac{\hbar \omega_0}{2 k_B T_1}\right) \right] + \right.$$
$$\left. + \frac{1}{(\omega_0 + \omega_s)} x_2^4 \left[ K_0(x_2) + K_2(x_2) \right] \cdot \left[ \coth\left(\frac{\hbar \omega_s}{2 k_B T_2}\right) + \coth\left(\frac{\hbar \omega_0}{2 k_B T_1}\right) \right] \right\}, \tag{10}$$

where $x_1 = 2|\omega_0 - \omega_s| z_0 / V$, $x_2 = 2(\omega_0 + \omega_s) z_0 / V$.

As it follows from (10), the tangential force is determined by two terms of which the first one can have both positive and negative sign, while the second one –only negative sign. $F_x > 0$ corresponds to acceleration and $F_x < 0$ – to deceleration of the particle. In the case $T_1 = T_2 = 0$ we have frictional force (the second term (10)) always, which also stems from eq.(5) if use is made of (8),(9). The corresponding exponential asymptotics of this quantum frictional force agrees with the results [1,2,8].

A necessary condition for the first addend in (10) to be positive implies

$$(\omega_s - \omega_0)\left[\coth\left(\frac{\hbar\omega_s}{2k_B T_2}\right) - \coth\left(\frac{\hbar\omega_0}{2k_B T_1}\right)\right] > 0 \tag{11}$$

This condition is fulfilled at $\omega_s > \omega_0, T_2 > T_1 = 0$ (hot surface and cold particle) and $\omega_s < \omega_0, T_1 > T_2 = 0$ (hot particle and cold surface). However, in order to get total accelerative force from (10), other parameters should be properly adjusted, as well. In general, this needs numerical calculation. One example of such an adjustment is given in the next section.

### 3. Numerical examples

First of all, we calculate frictional force for Na moving above the surface of Au assuming $T_1 = T_2 = 0$. The corresponding parameters are : $\alpha(0) = 24.08 \cdot 10^{-30} m^3$, $\omega_0 = 1.55 eV$ [32] and $\omega_P = 9 eV$ [26]. We will compare $F_x$ with attractive van der Waals force in the static case - $F_z$, which is given by

$$F_z = -\frac{3\hbar\alpha(0)\omega_0\omega_s}{8(\omega_0 + \omega_s)z_0^4}, \quad \omega_s = \omega_P/\sqrt{2} \tag{12}$$

Fig.1(a) shows fraction $F_x/F_z$ in dependence of distance $z_0$ and velocity $V$. We see that atom –surface frictional force can be as large as 1/3 of attractive van der Waals force. This is a striking result. Fig.1(b) shows the dependence $F_x$ on $z_0$ and $V$ in absolute units.

Strictly speaking, condition (11) can be valid for neutral atomic particles, too $(T_1 = 0, T_2 > 0, \omega_0 < \omega_s)$, but in order to get large enough values of accelerating force, one needs to assume the surface plasmon frequency $\omega_s$ to be close to $k_B T_2/\hbar$. This implies rather low frequencies $\omega_s < 0.2 eV$ even at $T_2 \sim 1000 \div 2000 K$. On the other hand, most of atomic transitions from the ground state are in the region $\omega_0 > 1 eV$, where (11) is not valid. Therefore, a possibility to get accelerating effect on ground state atoms seems to be of rather exotic.

However, a situation becomes more favorable in the case of nanoparticles moving above ionic dielectrics (like SiC, SiO2, etc.), where one finds intensive bands of light absorption in microwave frequency region. In order to demonstrate the accelerating effect, we have calculated $F_x$ in the case of MgO nanoparticle above a SiC surface. Both MgO and SiC materials can be described by similar one – oscillator model of dielectric function of the form

$$\varepsilon(\omega) = \varepsilon_\infty + \frac{(\varepsilon_0 - \varepsilon_\infty)\omega_T^2}{\omega_T^2 - \omega^2 - i\cdot\omega\gamma} \tag{13}$$

where $\varepsilon_0, \varepsilon_\infty$ correspond to $\omega = 0, \omega \to \infty$, $\omega_T$ is the transverse phonon frequency, $\gamma$ is the decrement factor. The numerical values of these parameters for MgO [13] and SiC [14] are listed in Table 1. We see from the given data that a difference between the characteristic frequencies of MgO and SiC is much larger than the corresponding $\gamma_1, \gamma_2$. This allows to use the resonance form of dielectric properties –eqs.(8),(9) with the proper modification, and finally –eq.(10). It can be simply shown that the corresponding values in (10) translate into the needed ones according to the following dictionary

$$\frac{\alpha(0)\omega_0\omega_s}{64} \qquad \omega_0 \qquad \omega_s$$

$$\frac{3R^3\omega_{T1}\omega_{T2}}{32}\frac{(\varepsilon_{10}-\varepsilon_{1\infty})(\varepsilon_{20}-\varepsilon_{2\infty})}{(\varepsilon_{1\infty}+2)^2(\varepsilon_{2\infty}+1)^2}\sqrt{\frac{\varepsilon_{1\infty}+2}{\varepsilon_{10}+2}}\sqrt{\frac{\varepsilon_{2\infty}+1}{\varepsilon_{20}+1}} \qquad \omega_{T1}\sqrt{\frac{\varepsilon_{10}+2}{\varepsilon_{1\infty}+2}} \qquad \omega_{T2}\sqrt{\frac{\varepsilon_{20}+1}{\varepsilon_{2\infty}+1}} \tag{14}$$

In relations (14) the lower indexes 1,2 numerate the particle (1) and surface (2), $R$ is the particle radius.

Fig.2 shows the velocity and temperature ($T_2$) dependence of tangential force at $R = 1nm$, $z_0 = 10nm$ and $T_1 = 0$ (cold particle and hot surface). We see that in the velocity interval $0.06 < V/V_B < 0.2$ ($V_B = 2.2\cdot 10^6 m/s$ is the Bohr velocity) the particle undergoes acceleration effect. Fig.3 demonstrates the corresponding distance and temperature dependence at $V/V_B = 0.1$, and Fig.4 –the distance and velocity dependence at fixed $T_2 = 1500K$. Note that melting points for MgO and SiC exceed $2500C$.

In order to judge about how large is accelerating force, one must compare the initial energy of the particle, $E_0 = MV^2/2$ with the power of acceleration, $P = F_x V$. Then for a MgO particle ($\rho = 3.58 g/cm^3, R = 1nm$) at $T_2 = 1500K$, $z_0 = 7nm$ and $V/V_B = 0.1$ we get $E_0 = 3.63\cdot 10^{-13} J$ and $P = 1.5\cdot 10^{-12} W$. This is also a very surprising outcome. It seems to be very interesting to study dynamics of such atomic clusters moving between the heated solid surfaces. As seen from Figs.2-4, the accelerating effect is localized in definite interval of velocities and distances. If a particle comes closer to the surface, or its velocity is too large, the tangential force becomes frictional, and the particle decelerates. On the other hand, far enough from the surface the particle can be accelerated once again.

## 4. Conclusions

We have analyzed behavior of tangential van der Waals force applied to a neutral atomic or nanoparticle moving with nonrelativistic velocity in close vicinity to a plane surface. Effects of temperature, velocity, distance and material parameters are studied. At zero temperatures of both the particle and surface, tangential force proves to be frictional –this is a manifestation of quantum friction. A surprising outcome is that for neutral atoms the corresponding value of the tangential force can reach 1/3 of the value of static Casimir –Polder force. In the case of small nanoparticles above a surface of solid, the tangential force can be accelerative at definite conditions. This is the second surprising outcome. Numerical calculations for Na atoms above Au surface and MgO nanoparticles above SiC surface demonstrate the expected magnitudes of frictional and accelerating forces and characteristic values of other parameters involved.

**References**

[1] Pendry J B 1997 *J.Phys.:Cond. Matter* **9** 10301

[2] Pendry J B, 2010 *New J.Phys.* **12** 0330208

[3] Dedkov G V, Kyasov A A 2002 *Phys. Solid State* **44/10** 1809 (a); 2003 *Phys. Low –Dim. Struct.* **1/2** 1(b)

[4] Hoye J S, Brevik I, 2010 *Europhys. Lett.* **91** 60003

[5] Hoye I S, I.Brevik I, 2010 arXiv: 1009.3135v1[quant-ph]

[6] Philbin T G, Leonhardt U, arXiv (2009) 0810.3750v3[quant –ph]

[7] Philbin T G, U.Leonhardt U, 2010 *New J.Phys.* **12** 068001

[8] Barton G, 2010 *New J.Phys.* **12** 113045

[9] Dedkov G V, Kyasov A A 2010, *Surf. Sci.* **604** 561; 2010 *Phys. Solid State* **52/2** 409; 2009 *Nanostruct.Math.Phys.Model.* **1/2** 5 (in Russian)

[10] Volokitin A I, Persson B N J 2008, *Phys. Rev.* **B78** 155437

[11] Volokitin A I, Persson B N J 2007, *Rev.Mod.Phys.* **79** 1291

[12] Bruhl R, Fouquet P, Grisenti R E, Toennis J P, Hegerfeldt G C, Kohler T, Stoll M, and Walter C 2002, *Europhys. Lett.* **59** 357

[13] Blakemore J S 1982 , *J.Appl.Phys.* **53** R123

[14] Palik D E (ed) 1985 *Handbook of Optical Constants of Solids,* (New York: Academic)

Table 1

Parameters of ionic dielectrics

| Material | $\varepsilon_0$ | $\varepsilon_\infty$ | $\omega_T, eV$ | $\gamma, eV$ |
|---|---|---|---|---|
| MgO | 9.8 | 2.95 | 0.049 | $4.9 \cdot 10^{-4}$ |
| SiC | 9.8 | 6.7 | 0.098 | $5.9 \cdot 10^{-4}$ |

FIGURE CAPTIONS

Fig.1(a,b) Tangential forces on atoms of Na above Au substrate. (a) $F_x / F_z$ ; (b) $-F_x$ expressed in pN. Lines 1,2,3,4 correspond to velocities $V/V_B = 2, 5, 10, 20$.

Fig.2 The velocity and temperature dependence of tangential forces on a MgO particle ($R = 1nm$) above a SiC substrate. The solid, dotted, dashed and dashed –dotted lines correspond to the substrate temperatures of 1500, 1200, 900 and 600K. A particle has zero temperature, the distance to surface is $10nm$.

Fig.3 The distance and temperature dependence of tangential force on a MgO particle ($R = 1nm$) above a SiC substrate. Solid, dotted, dashed and dashed –dotted lines correspond to the substrate temperatures of 1500, 1200, 900 and 600K, $V/V_B = 0.1$.

Fig.4 The distance and velocity dependence of tangential force on a MgO particle ($R = 1nm$) above a SiC substrate with temperature 1500K. Solid, dotted, dashed and dashed –dotted lines correspond to the particle velocities of 0.02, 0.04, 0.08 and 0.16 Bohr units.

FIGURE 1(a,b)

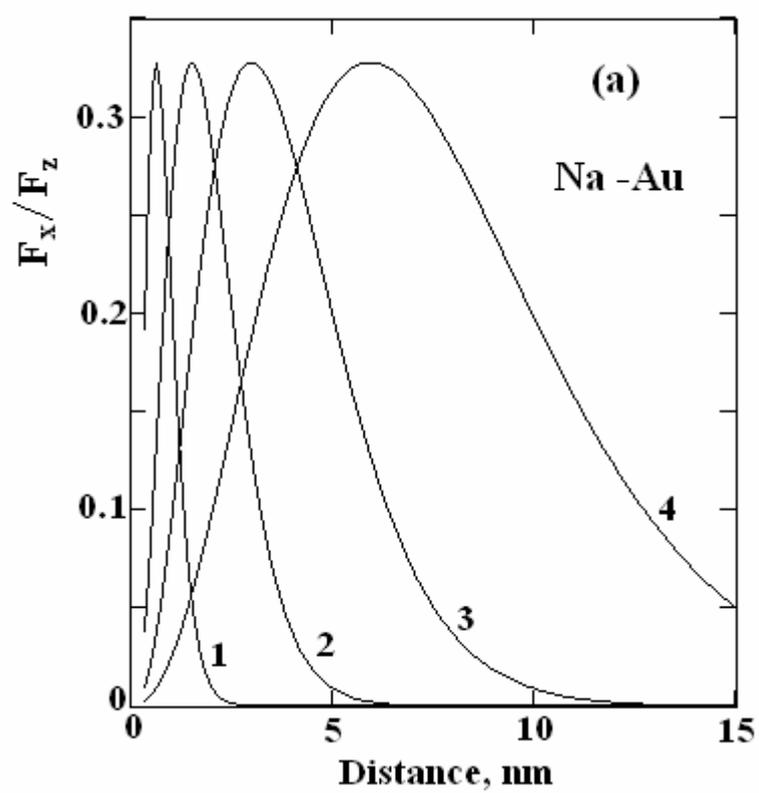

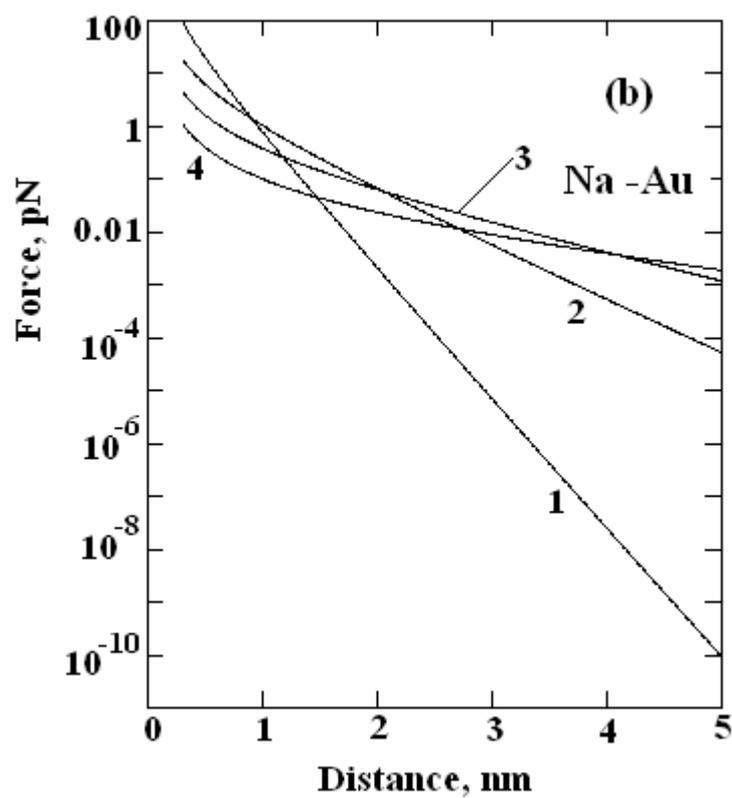

FIGURE 2

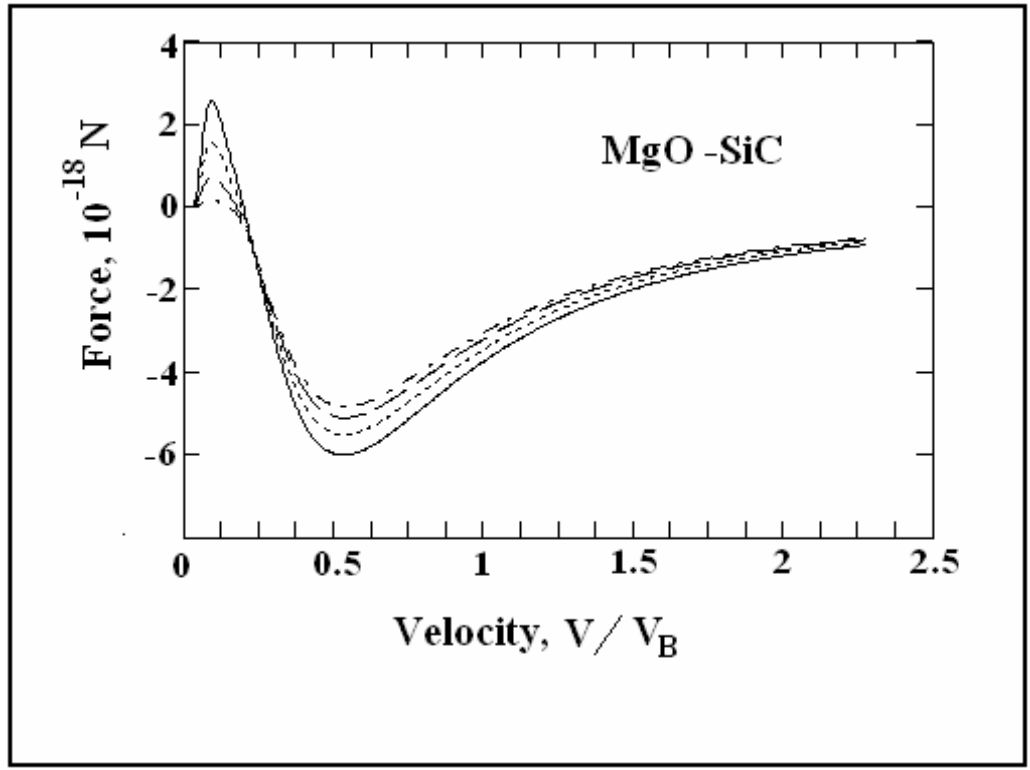

FIGURE 3

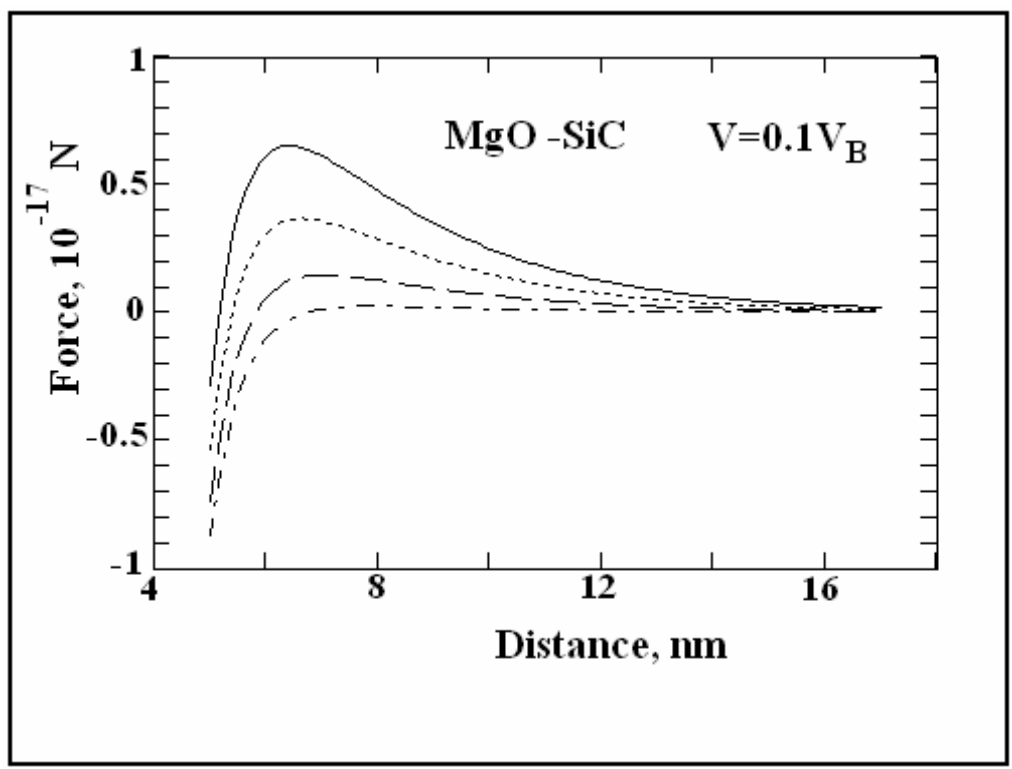

FIGURE 4

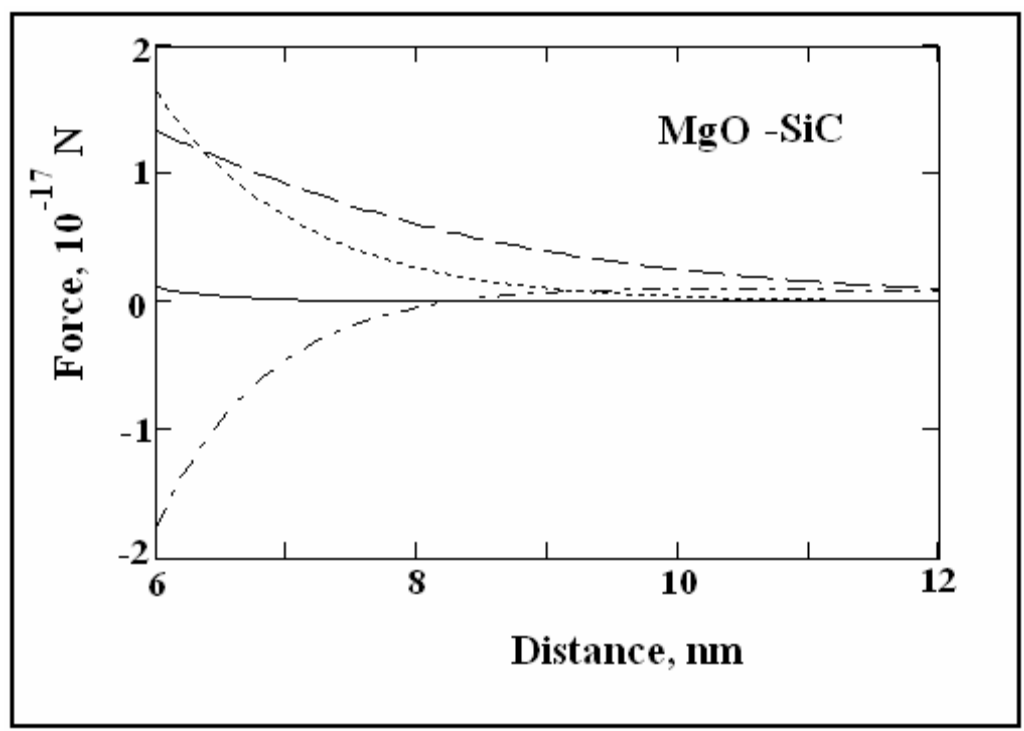